%% file: main.tex
\documentclass[sigconf, screen]{acmart}
\usepackage[colorinlistoftodos]{todonotes}
\usepackage{pifont}
\usepackage[skip=1pt]{caption}
\usepackage{enumitem}
\usepackage{makecell}
\usepackage{balance}
\usepackage{multirow}
\newcommand\blfootnote[1]{%
  \begingroup
  \renewcommand\thefootnote{}\footnote{#1}%
  \addtocounter{footnote}{-1}%
  \endgroup
}

\setcopyright{none} 
\acmConference[Anonymous Submission to ACM CCS 2019]{ACM Conference on Computer and Communications Security}{Due 15 May 2019}{London, TBD}
\acmYear{2019}

\begin{document}
\title{Garbled EDA: Privacy Preserving Electronic Design Automation}

\author{Mohammad Hashemi}
\email{mhashemi@wpi.edu}
\affiliation{\institution{Worcester Polytechnic Institute}
\city{Worcester, MA, USA}}
\author{Steffi Roy}
\email{steffiroy@ufl.edu}
\affiliation{\institution{University of Florida}
\city{Gainesville, FL, USA}}
\author{Fatemeh Ganji}
\email{fganji@wpi.edu}
\affiliation{\institution{Worcester Polytechnic Institute}
\city{Worcester, MA, USA}}
\author{Domenic Forte}
\email{dforte@ece.ufl.edu}
\affiliation{\institution{University of Florida}
\city{Gainesville, FL, USA}}


\begin{abstract}

The complexity of modern integrated circuits (ICs) necessitates collaboration between multiple distrusting parties, including third-party intellectual property (3PIP) vendors, design houses, CAD/EDA tool vendors, and foundries, which jeopardizes confidentiality and integrity of each party's IP. 
IP protection standards and the existing techniques proposed by researchers are ad hoc and vulnerable to numerous structural, functional, and/or side-channel attacks.
Our framework, Garbled EDA, proposes an alternative direction through formulating the problem in a secure multi-party computation setting, where the privacy of IPs, CAD tools, and process design kits (PDKs) is maintained. 
As a proof-of-concept, Garbled EDA is evaluated in the context of simulation, where multiple IP description formats (Verilog, C, S) are supported. 
Our results demonstrate a reasonable logical-resource cost and negligible memory overhead. 
To further reduce the overhead, we present another efficient implementation methodology, feasible when the resource utilization is a bottleneck, but the communication between two parties is not restricted. 
Interestingly, this implementation is private and secure even in the presence of malicious adversaries attempting to, e.g., gain access to PDKs or in-house IPs of the CAD tool providers. \blfootnote{\copyright ACM. This is the author version of the paper accepted for presentation at \textbf{2022 IEEE/ACM International Conference On Computer Aided Design}.  Personal use of this material is permitted. Permission from ACM must be obtained for all other uses, in any current or future media, including reprinting/republishing this material for advertising or promotional purposes, creating new collective works, for resale or redistribution to servers or lists, or reuse of any copyrighted component of this work in other works.}
\end{abstract}


\maketitle

\input{Introduction_Edited}
\input{Background}
\input{Garbled_Compiler}
\input{Garbled_Compiler_Evaluation}
\input{Conclusion}

\section{Acknowledgements}
This work was supported by Semiconductor Research Corporation (SRC) under Task IDs 2991.001 and 2992.001.

\bibliographystyle{ACM-Reference-Format}
\bibliography{references}

\end{document}

%% file: Introduction_Edited.tex
\section{Introduction}\label{sec:introduction}
Nowadays, the electronics supply chain is distributed worldwide,  involving untrusted parties in design and manufacturing.
Intellectual property~(IP) piracy and tampering are prominent examples of the security risks introduced by such outsourcing and globalization.
As a collection of reusable design specifications (e.g., RTL, netlists, etc.), semiconductor IPs are created and owned by one party, but are often licensed to other parties. 
To accelerate the integrated circuit (IC) design process, third-party IPs (3PIPs) are procured in the forms of soft, firm, and hard IPs cf.~\cite{chhotaray2017standardizing}. 
At a design house, IC designers integrate 3PIPs with their own in-house IPs. 
During the integration process, however, the parties may not trust each other since, driven by economic incentives, some may attempt to extract, re-use, overuse, or modify design information from the IPs in an unauthorized manner. 
While the industry relies on non-disclosure and other agreements as legal disincentives, they are not insurmountable technical barriers.

One category of solutions devised to actively prevent such attacks  includes logic locking and encryption-based techniques, which incorporate locking mechanisms to guarantee proper IC/IP functionality only in the presence of a correct key~\cite{dupuis2019logic}. 
Logic locking techniques have been proposed to protect the IP at almost all stages of hardware design; however, various oracle and oracle-less attacks have rendered each new logic locking useless by successfully extracting the key~\cite{zamiri2019threats}. 
Other IP protection methods have applied encryption/decryption, see, e.g.,~\cite{roy2010ending,guin2016fortis}. 
For this, to remotely unlock ICs and IPs, an on-chip root-of-trust executing protocols between IPs and the IP-owner is required. 
Besides the flaws in the protocols~\cite{maes2009analysis} and on-chip root-of-
trust requirement, they could also be vulnerable to side-channel analysis targeting commonly applied encryption/decryption cores, e.g., RSA~\cite{roy2010ending}. 
Even a combination of locking- and encryption-based countermeasures, e.g., the proposal in~\cite{stern2021aced}, has the same shortcoming as previous IP protection methods. 

As a remedy for problems stemming from on-chip solutions, the role of a trusted entity, which can unlock the IP, is often played by electronic design automation (EDA) tools. Conventionally, EDA tools provide powerful solutions for simulation, verification, testing, and power, performance, and area (PPA) optimization.   
Although EDA tools have often been assumed to be trusted, the security and privacy of IPs has not been involved in their specifications. 
In an attempt to provide such capabilities, the IEEE P1735 standard~\cite{ieee_p1735} for encryption and management of IPs was created. 
Unfortunately, it was demonstrated that this standard suffers from  side-channel leakage by EDA tools, and accordingly, to the disclosure and undetectable tampering of the IPs~\cite{chhotaray2017standardizing,ieee_p1735_broken}.  
In addition, the private keys relied upon by the standard may be stolen directly from the EDA tool source code/binaries, as recently demonstrated in~\cite{industry_wide}.
Such attacks highlight the importance of EDA tools in protecting the confidentiality of IPs. 
In fact, a unified, comprehensive security-centric EDA flow should be made feasible, where security-relevant metrics and \textit{provably secure} mechanisms are taken into account~\cite{knechtel2020towards}.

To address this, we introduce Garbled EDA, a framework that represents a fundamental paradigm shift, but nevertheless can easily work in conjunction with commercial EDA tools (see Table~\ref{Tab_3} for a comparison between the approaches). 
Importantly, Garbled EDA considers a \emph{zero trust} environment, where three parties (3PIP owner, IC designer, and EDA tool vendor) involved in the process of IC design are mutually distrusting. 
Specifically, we discuss how secure function evaluation (SFE) and private function evaluation (PFE) schemes can be adapted to the specific needs of a secure and privacy preserving EDA tool. 
Our initial goal is to enable the user (IC designer) to obtain the technology mapped/optimized circuit or a simulation of it without disclosing the information about the process design kit (PDK), IP, or EDA binaries to other parties.  

\vspace{0.5ex}\noindent\textbf{Contributions: }
Our contributions are summarized below. 
\vspace{-0.5ex}
\begin{itemize}[leftmargin=*]
    \item \noindent We propose Garbled EDA, a framework that unlike logic locking and related approaches relies on \textit{provable security} to protect the IC design process. 
    Note that Garbled EDA can handle common EDA tools/steps such as formal verification, DFT, ATPG, etc., as well as emerging security analysis tools. 
    Nevertheless, we focus our discussions on compilation and simulation for simplicity.
  
    \item \noindent We utilize open-source tools in our framework, so that IP vendors can easily incorporate them to secure their IPs in hostile environments. Moreover, Garbled EDA prevents the reveal of any information about the given IP while simulating, synthesizing, and generating the Garbled EDA output. 
    This is also guaranteed even under scenarios where IC designers would attempt to actively tamper with the IPs (the so-called malicious adversary). 
    \item \noindent We evaluate the output of Garbled EDA using Xilinx Vivado 2021 as well as on an Artix-7 FPGA. As proof-of-concept, we present an IP-specific simulator as well as a selector-based technique to alleviate the overhead of generating simulators for multiple IPs to be integrated into the same IC design. 
    This reduces the logical resource utilization costs of Garbled EDA by up to five times. Further, we examine the resource-performance tradeoffs for a Garbled EDA simulator implementation flow with an improved hardware resource efficiency.
    \item \noindent We compare the implementation costs between MIPS and ARM architecture-based IP-specific simulators, which provide the IP owners an insight to choose between them based on the application resource limitations.
    \end{itemize}


\noindent\textbf{Organization: }An overview of this paper is as follows. Section~\ref{sec:back}  introduces our adversary model and SFE/PFE protocols. The Garbled EDA framework is presented in Section~\ref{sec:Garbled_Simulator}, whereas our evaluation results are presented in Section~\ref{sec:Garbled_Simulator_Evaluation}. Section~\ref{sec:conclution} concludes the paper.


\begin{table}[t!]
\scriptsize
    \setlength{\tabcolsep}{0.5em}
    \renewcommand{\arraystretch}{1.5}
        \begin{center}
        \caption{Garbled EDA vs. existing methods. \label{Tab_3}}
            \begin{tabular}{|c|c|c|c|}
                \hline
                \textbf{Problems}&\textbf{IEEE P1735}&\textbf{SFE-based EDA}&\textbf{Garbled (PFE) EDA}\\ 
                
                \hline
                \makecell{Supports privacy of designer\\ inputs (e.g., PDK)}&\textcolor{red}{\ding{55}}&\textcolor{green}{\checkmark}&\textcolor{green}{\checkmark}\\ 
                \hline
                Safe from CAD/EDA tool hacking&\textcolor{red}{\ding{55}}&\textcolor{green}{\checkmark}&\textcolor{green}{\checkmark}\\ 
                \hline
                Handles untrusted CAD/EDA vendor&\textcolor{red}{\ding{55}}& \textcolor{red}{\ding{55}}/\textcolor{green}{\checkmark}&\textcolor{green}{\checkmark}\\
                \hline
                Supports CAD/EDA tool privacy &\textcolor{red}{\ding{55}}&\textcolor{red}{\ding{55}}&\textcolor{green}{\checkmark}\\
                \hline
            \end{tabular}
    \end{center}
   \vspace{-10pt}
\end{table}

%% file: Background.tex
\section{Background Information} \label{sec:back}

\vspace{-2pt}

\begin{figure}[t!]
\centering \noindent
\includegraphics[width=0.95\columnwidth]{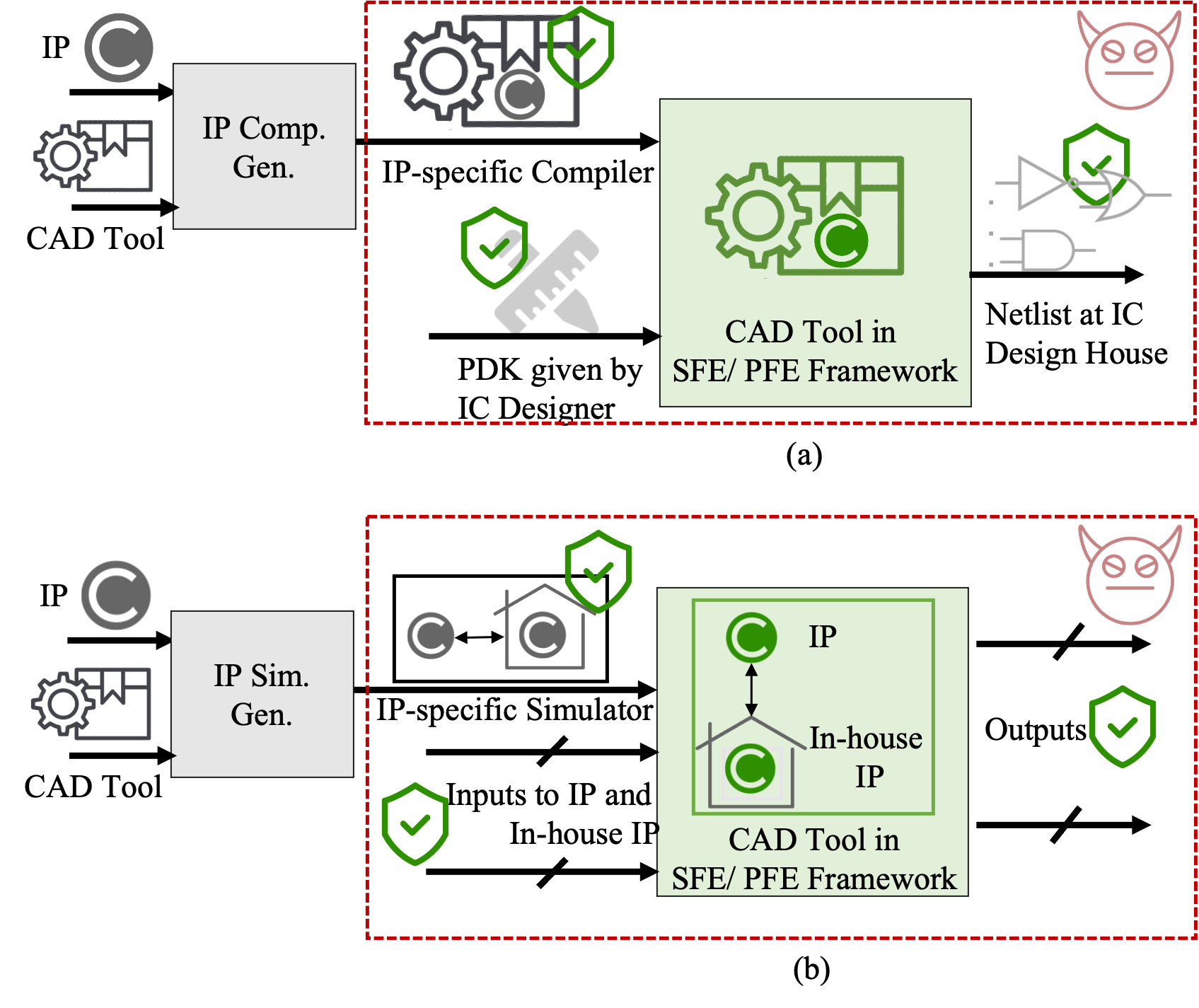}
\caption {Proposed CAD/EDA compilation/simulation of IP under various scenarios: (a) compilation in a secure and private way, where IP and PDK inputs are secure, and the IP-specific compiler can be private.   
The output netlist of the compilation can also be made private;   (b) inputs given to IPs can be made secure, and the IP-specific simulator can be private. 
The simulator's output can also be made private. 
The adversary at the design house could be either honest-but-curious or malicious. 
In the malicious case, the IC designer may tamper with the IP-specific compiler/simulator to extract the IP. 
The proposed approach can deal with both of these types of adversaries. 
\vspace{-14pt}}
\label{fig:sim_comp_sfe}
\end{figure}

\subsection{Adversary Model}
Classically, two main categories of adversaries have been considered in the SFE/PFE context. A \textit{semi-honest} (so-called honest-but-curious or passive) adversary is expected to execute a protocol correctly and not deviate from the protocol specification, but may attempt to learn more from the information at hand. 
All three main parties in compilation and simulation scenarios -- 3PIP owner, IC designer, and CAD/EDA tool vendor -- may act as a semi-honest adversary. 
The CAD tool source code can be targeted and extracted by the IC designer and/or the IP owner\footnote{There are examples of proprietary software being hacked for unauthorized use.}. 
The IP owner might also be interested in accessing the process design kit (PDK), e.g., the properties of standard cells applied by the IC designer, while the IC designer may attempt to extract the raw IP, i.e., in RTL or higher-level language. 
An example of measures devised to thwart the latter attack is IEEE P1735~\cite{ieee_p1735}, whose shortcomings were discussed in Section~\ref{sec:introduction}. The second type of adversaries constitutes \textit{malicious} (so-called active) ones that may arbitrarily deviate from the protocol execution and attempt to cheat. 
The honest-but-curious adversary is often taken into account since it serves as a basis for proving the security even in the presence of a malicious adversary (see Section~\ref{sec:compiler_foundation} for a discussion on this). 
Nevertheless, in this paper, we consider both types. 
Concretely, the IC designer can act as a malicious adversary and attempt to manipulate the inputs of the IP owner arbitrarily in the hope of extracting information about the IPs (see Section~\ref{sec:prelim_background} for more details). 

To fulfill these security-critical goals, our methodology relies on SFE/PFE protocols, where the inputs of the parties remain private during the execution of the protocol thanks to the \textit{provable security} of SFE schemes. 
Moreover, if we consider the input of the EDA tool vendor as its private input, the PFE techniques enable us to protect not only the inputs, but also the EDA tool against the adversaries in both compilation and simulation cases
, see Figure~\ref{fig:sim_comp_sfe}. 
When performing simulation, the IC designer might apply proprietary or export-controlled inputs as well as in-house IPs that should be protected against adversaries, e.g., the EDA tool vendor. 
This can be obtained as a byproduct of executing an SFE/PFE protocol in the case of simulation. 
However, if our interactive framework is employed to securely and privately use security analysis tools, inputs and outputs given by the IC designer should be definitely kept secure. 
In the same vein, after employing a CAD/EDA tool for synthesis, the gate-level netlist could be an asset to be protected from the IP owner and CAD tool vendor, who may derive information in an unauthorized way, see Figure~\ref{fig:sim_comp_sfe}. 
Although IC designers perform this process at their facilities, if the output is generated interactively (e.g., in the context of SFE protocols), the security of the output should be guaranteed. 
The building blocks of SFE/PFE schemes to achieve these goals are formally introduced below. 


\subsection{Background on SFE/PFE Protocols} \label{sec:prelim_background}

\begin{figure}[t!]
\centering \noindent
\includegraphics[width=0.9\columnwidth]{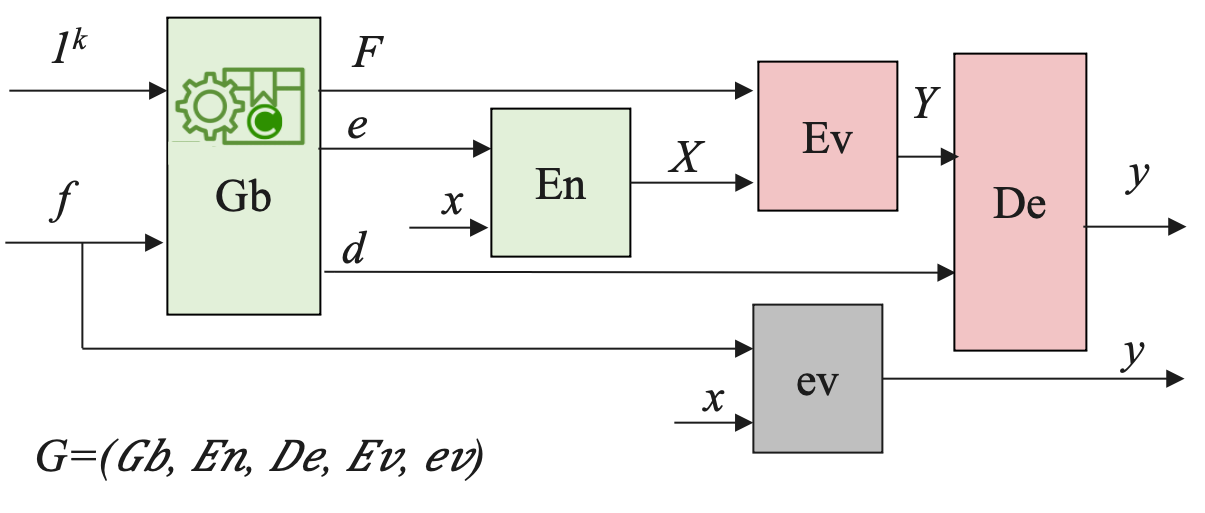}
\caption {A generic garbling scheme $G = (Gb, En, De, Ev, ev)$ cf.~\cite{bellare2012foundations}. Our proposed secure and private EDA is built upon $G$. 
For Garbled EDA, the blocks in green show the operations performed by the CAD tool vendor, whereas the red ones indicate the IC designer's operations. 
\vspace{-10pt} }
\label{fig:garbling_scheme}
\end{figure}

SFE protocols enable a group of participants to compute the correct output of some agreed-upon function $f$ applied to their private inputs without revealing anything else. 
One of the commonly-applied SFE protocols is Yao's garbled circuit~\cite{yao1986generate}, a two-party computation protocol. To formalize this protocol, we employ the notions and definitions provided in~\cite{bellare2012foundations} to support modular and simple but effective analyses. 
In this regard, a garbling algorithm $Gb$ is a randomized algorithm, i.e., involves a degree of randomness. 
$Gb(f)$ is a triple of functions $(F, e, d) \leftarrow Gb(f)$ that accepts the function $f: \{0, 1\}^n \rightarrow \{0, 1\}^m$ and the security parameter $k$. 
$Gb(f)$ exhibits the following properties. 
The encoding function $e$ converts an initial input $x \in \{0, 1\}^n$ into a garbled input $X = e(x)$, which is given to the function $F$ to generate the garbled output $Y = F(X)$. 
In this regard, $e$ encodes a list of tokens (so-called labels), i.e., one pair for each bit in $x \in \{0, 1\}^n$:  $En(e, \cdot)$ uses the bits of $x = x_1 \cdots x_n$ to select from $e = (X^0_1 ,X^1_1, \cdots ,X^0_n,X^1_n)$ and obtain the sub-vector $X = X^{x_1}_1,\cdots,X^{x_n}_n$.
By reversing this process, the decoding function $d$ generates the final output $y = d(Y)$, which must be equal to $f(x)$. 
In other words, $f$ is a combination of probabilistic functions $d\circ F\circ e$. 
More precisely, the garbling scheme $G=(Gb, En, De, Ev, ev)$ is composed of five algorithms as shown in Figure~\ref{fig:garbling_scheme}, where the strings $d$, $e$, $f$, and $F$ are used by the functions $De$, $En$, $ev$, and $Ev$. 

\begin{figure}[t]
\centering \noindent
\includegraphics[width=0.8\columnwidth]{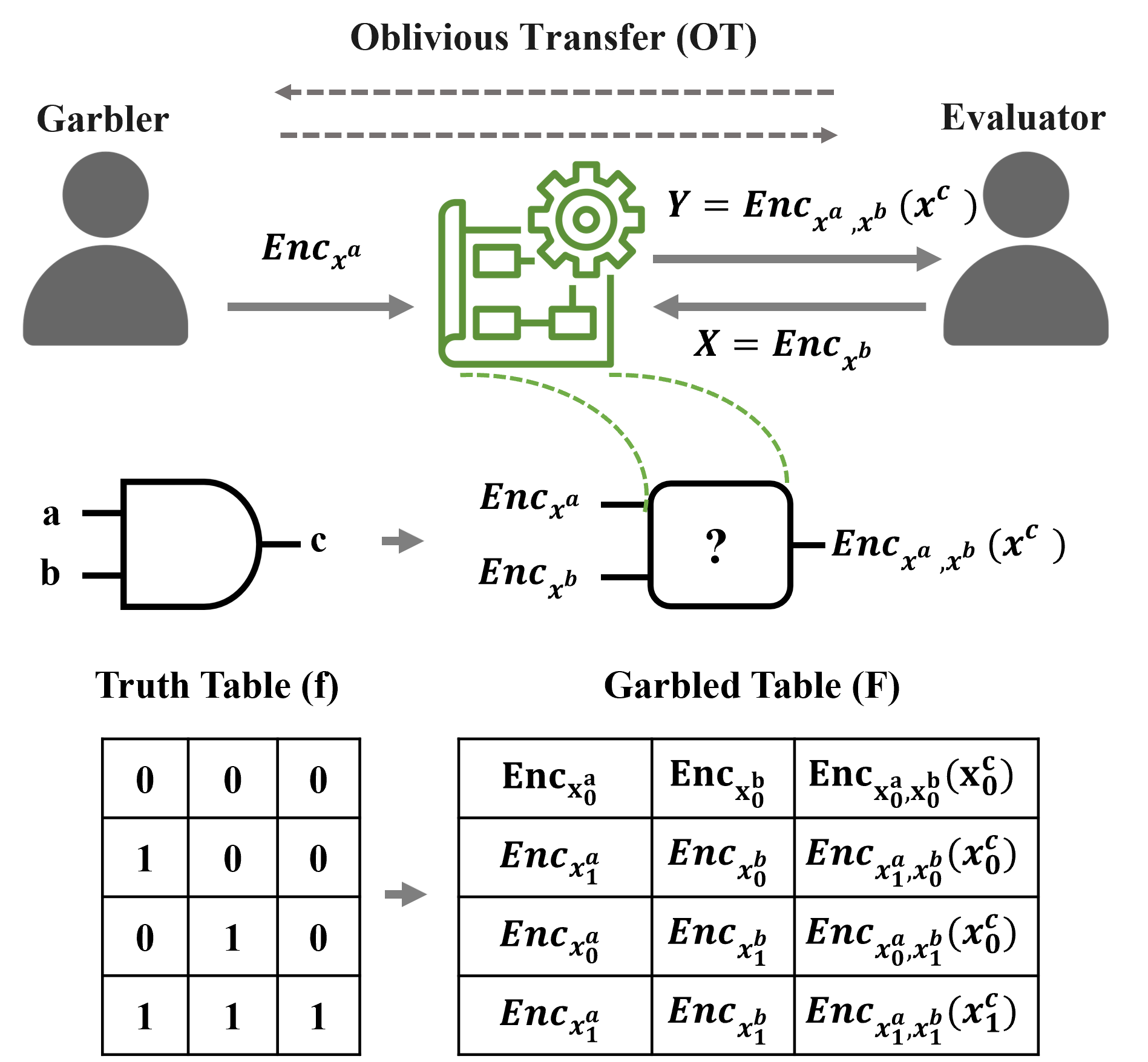}
\caption {An example of AND gate garbling and evaluation using SFE/PFE protocol.
\vspace{-10pt}}
\label{fig:Garbling_AND}
\end{figure}

\vspace{5pt}\noindent\textbf{Security of garbling schemes: } 
For a given scheme, the security can be roughly defined as the impossibility of acquiring any information beyond the final output $y$ if the party has access to $(F,X, d)$. 
Formally, this notion is explained by defining the side-information function $\varPhi(\cdot)$. 
Based on the definition of this function, an adversary cannot extract any information besides $y$ and $\varPhi(f)$ when the tuple $(F,X, d)$ is accessible. 
As an example of how the function $\varPhi(\cdot)$ is determined, note that for an SFE protocol, where the privacy of the function $f$ is not ensured, $\varPhi(f)=f$. 
Thus, the only thing that leaks is the function itself.
On the other hand, when a PFE protocol is run, $\varPhi(f)$ is the circuit/function's size, e.g., number of gates.

\vspace{5pt}\noindent\textbf{Oblivious transfer (OT):} 
This is a two-party protocol where party~2 transfers some information to party~1 (so-called Evaluator); however, party~2 remains oblivious as to what information party~1 actually obtains. 
A form of OT widely used in various applications is known as ``chosen one-out-of-two'', denoted by  1-out-of-2 OT. 
In this case, party~2 has bits $X^0$ and $X^1$, and party~1 uses one private input bit $s$. 
After running the protocol, party~1 only gets the bit $X^s$, whereas party~2 does not obtain any information on the value of $s$, i.e., party~2 does not know which bit has been selected by party~1. 
This protocol can be extended to support the $n$-bit case, where party~1 bits $x_1, \cdots, x_n$ are applied to the input of party~2 $X^0_1,X^1_1, \cdots, X^0_n,X^1_n$ to obtain $X^{x_1}_1,\cdots,X^{x_n}_n$. 
This is possible by sequential repetition of the basic protocol~\cite{bellare2012foundations}. 
It has been proven that 1-out-of-2 OT is universal for 2-party SFE, i.e., OT schemes can be the main building block of SFE protocols~\cite{kilian1988founding}.  

\vspace{5pt}\noindent\textbf{Building a PFE protocol:} To construct a PFE protocol from the scheme $G$, we first define a  polynomial algorithm $\Pi$ that accepts the security parameter $k$ and the (private) input of the party~\cite{bellare2012foundations}. 
The PFE scheme is a pair $\mathcal{F} = (\Pi, ev)$, where $ev$ is as defined for the garbling scheme (see Section~\ref{sec:garbled_eda_implementation} for more information about realization of $\Pi$). 
The scheme $\mathcal{F}$ enables us to securely compute the \emph{class} of functions $\{ev(f, \cdot) : f \in \{0, 1\}^*\}$, i.e., any function that $G$ can garble. 
The security of the PFE scheme $\mathcal{F}$ relies on the security of the SFE protocol underlying $\mathcal{F}$; however, $\varPhi(f)$ is the circuit size, i.e., the function $f$ remains private when executing the SFE protocol. 

\vspace{5pt}\noindent\textbf{Efficient tools for garbling circuits:} 
TinyGarble and its variant TinyGarble2 are garbling frameworks that support Yao's protocol and use hardware-synthesis tools to generate circuits for secure computation automatically. 
For this, party~2 possessing a Boolean circuit $f$ performs garbling by encrypting the truth table of each gate $F$, see Figure~\ref{fig:garbling_scheme}. 
Inputs to these gates are the encrypted inputs of party~2 and intermediate gates' outputs ($d$ and $e$).  
Afterward, the encrypted input labels and the truth tables of all gates are sent to party~1, who evaluates the garbled circuit ($Ev$) and decodes the output ($De$) by executing the OT between parties.
Figure~\ref{fig:Garbling_AND} illustrates an example of an AND evaluation using SFE/PFE protocol.
Based on the AND logic truth table ($f$), Garbler (party~2) generates Garbled tables ($F$), input encryption labels ($e=Enc_{X^b}$), output decryption labels ($y=d(Enc_{x^a,x^b}(x^c))$).
The tuple $(F,e,d)$ corresponds to the function $Gb$ in Figure~\ref{fig:garbling_scheme}.

The main advantage of TinyGarble and TinyGarble2 is their scalability enabled  by exploiting a sequential circuit description for garbled circuits and garbling optimization techniques such as Free-XOR \cite{kolesnikov2008improved}, Row Reduction \cite{naor1999privacy}, and Garbling with a Fixed-key Block Cipher.
A custom circuit description (SCD) allows TinyGarble to evaluate the Boolean circuit using the 2-party computation protocol securely. 
In addition to this, TinyGarble2 has introduced a practical method to deal with the case of \emph{malicious} adversaries through the adoption of provably secure protocols given in~\cite{lindell2016fast}. 

%% file: Garbled_Compiler.tex
\section{Garbled EDA}\label{sec:Garbled_Simulator}

\subsection{Foundations of Secure and Private EDA}\label{sec:compiler_foundation}
\noindent\textbf{Protocol flow:}
Here we provide insight into how SFE/PFE schemes can be tailored to the needs of a secure EDA compiler/simulator. 
The general flow of our secure and private EDA tool is illustrated in Figure~\ref{fig:garbling_scheme}. 
The process is represented by evaluating a function $f$ against some inputs $x$ to obtain the output $y$, where this process is run in the context of a garbling protocol $G$. 
For the sake of simplicity, suppose that the 3PIP owner trusts the EDA tool vendor so that the raw IP is included in the source code of the tool and the garbled IP-specific tool, hereafter called Garbled EDA, is provided to the IC designer for compilation/simulation (the extension to the case with untrustworthy vendor discussed next). 
Thus, in our proposed approach, the designer is never in possession of the raw EDA binaries or raw IP unlike the typical EDA flow today. 
Rather, the garbled tables represent a private, IP-specific compiler/simulator that the designer uses (see Figures~\ref{fig:sim_comp_sfe}(a--b)). 

Let $f=f_{CAD}\circ f_{IP}$ denote the IP-specific compiler/simulator function, which combines the source code of the EDA with the raw IP. 
The IC designer aims to obtain the technology mapped/optimized circuit or simulation output $y$ without disclosing the information about the PDK or its simulation inputs. 
To achieve this, here we give an example of SFE protocol $G$ that has OT at its core and follows Yao's garbling principle, i.e., the garbling protocol $G=(Gb, En, De, Ev, ev)$ as shown in Figure~\ref{fig:garbling_scheme}. 
To execute the protocol, party~2 (CAD tool vendor) conducts $(F, e, d) \leftarrow Gb(1^k,f)$ on inputs $1^k$ and $f$ and parses $(X^1_0 ,X^1_1, \cdots ,X^0_n,X^1_n) \leftarrow e$. 
Afterward, the CAD tool vendor sends $F$ and (optionally) $d$ to the IC designer. 
In order to perform the function $Ev$, the IC designer and the CAD tool vendor run the OT, where the former has the selection string $x$ and the latter party has already parsed $(X^1_0 ,X^1_1, \cdots ,X^0_n,X^1_n)$. 
Hence, the IC designer can obtain $X=X^{x_1}_1,\cdots,X^{x_n}_n$ and consequently, $y \leftarrow De(d, Ev(F,X))$. 
Note that even with the tuple $(F,X, d)$ in hand, the IC designer acting as an adversary cannot extract any information besides $y$ and $\varPhi(f)$.  
Moreover, although the EDA tool provider has access to $(F,e,d)$, no information on $x$ leaks. 
In a compilation scenario, $x$ represents the PDK, whereas in a simulation case, $x$ denotes the inputs given by the IC designer. 
Nevertheless, if $G$ is an SFE scheme, $\varPhi(f)=f$; hence, $f$ (i.e., the combination of EDA tool code and IP) is revealed to the IC designer. 
To keep the CAD tool and IP private, we build a PFE/SFE protocol as described in Section~\ref{}. 

Under a \emph{malicious} adversary scenario, the IC designer receiving the function $F$ and encoding string $e$ corresponding to garbled tables and labels attempts to extract the information regarding the IP. 
This can be done by actively tampering with the garbled tables, observing the output $y$ to reveal $e$, and decrypting the relevant row of the garbled table. 
To overcome this issue, the concept of cut-and-choose, see, e.g.,~\cite{lindell2007efficient,lindell2016fast}, has been widely employed. 
The cut-and-choose protocol suggests that the garbler constructs many garbled circuits instead of one and sends them to the evaluator.
This principle is interpreted and employed in a different and more efficient way in~\cite{hussain2020tinygarble2}, where one OT is performed per gate, as opposed to per input bit in the honest-but-curious model. 
Hence, the malicious adversary cannot effectively decrypt all the table rows by relating the output $y$ to the output of each manipulated table. 

Another interesting characteristic of SFE/PFE schemes is their ability to adapt to specific scenarios, where the output $y$ should also be protected. 
In our case, if the decryption of $Y$ is performed in a hostile environment, the privacy of the gate-level netlist can easily be preserved by applying a one-time message authentication code (MAC) to the output and XORing the result with a random input to hide the outcome. 
These operations are included in the IP-specific-compiler/simulator and naturally increase its size and the input fed by the IP owner; however, the increase is linear in the number of output bits and considered inexpensive~\cite{lindell2007efficient}. 

Next, we discuss some properties of the protocol $G$ for the purpose of secure and private EDA. 
Specifically, we should deal with three main aspects of the design of such compilers/simulators: (1) number of parties involved in the computation, (2) number of inputs to be evaluated securely per garbled compiler, and (3) interactive and non-interactive scenarios.

\vspace{5pt}\noindent\textbf{Multiparty garbling: }
Despite the fact the Yao's pioneering SFE protocol has been formulated for the two-party case, other protocols can be applied in a multiparty scenario~\cite{goldreich2019play}. 
Hence, in our framework, each of the parties (3PIP owner, IC designer, and EDA tool vendor) can communicate through a multiparty SFE protocol. 
In doing so, the IP owner gives the function $f_{IP}$ as the input. 
Since the inputs of the users are guaranteed to be secure, the privacy of the protocol is further fulfilled. 
Nevertheless, to demonstrate proof-of-concept using available tools, we consider the case, where an IP owner trusts the EDA tool vendor so that the raw IP is included in the source code of the IP-specific tool. 

\vspace{5pt}\noindent\textbf{Evaluation of the Garbled EDA tool against multiple inputs: }
When compiling a design, the IC designer may interact with the protocol through the extension of OT. 
It could be feasible to send the entire set of inputs at once (sometimes called parallel execution); however, sequential execution of the OT and the SFE protocol built upon that leads to a more resource-efficient implementation since all the labels should not be stored together. 
Note that to guarantee security, a fresh garbled function $F$ for each IC designer/a set of inputs should be generated. 
This has important implications: (1) the garbled function $F$ is specific to a given function $f$, i.e., it is not possible to use it to execute the protocol for other functions (IPs and CAD tools), and (2) the protocol $G$ can be extended to support multiple executions. 
For this purpose, a straightforward approach is to run a maliciously-secure protocol multiple times~\cite{huang2014amortizing}. 
This protocol relies on the cut-and-choose protocol~\cite{lindell2016fast}, which is considered for malicious case in our study (see Section~\ref{sec:garbled_eda_implementation}).  


\vspace{5pt}\noindent\textbf{Interactive and non-interactive garbling: }
If an SFE protocol with garbling scheme $G$ is built upon an OT, after receiving the inputs $F$ and $d$, party~1 can initiate the OT protocol to obtain the output $y$. In other words, the interaction between parties happens solely when the OT protocol is executed. 
Therefore, if a non-interactive OT can be realized, the SFE protocol can be run non-interactively. 
This has been achieved by incorporating appropriate hardware such as trusted platform modules (TPMs)~\cite{gunupudi2008generalized} and simple tamper-resistant hardware~\cite{goldwasser2008one}. 
These primitives can also offer an important byproduct that is one-time or limited-time execution, i.e., the compilation/simulation can occur only once or for a limited time, which can be helpful for enforcing IP/EDA license agreements.

\subsection{Implementation of Garbled EDA}\label{sec:garbled_eda_implementation}
In this section, first, a high-level description of the Garbled EDA approach has been given; then, the simulator implementation flow with the maximum performance has been explained; and finally, the flow of the simulator implementation with hardware resource efficiency is proposed.

\vspace{5pt}\noindent\textbf{Garbled EDA high level description: }
When defining the PFE scheme $\mathcal{F}$, it is mentioned that $\mathcal{F}$ can securely and privately compute \emph{any} function, which can be garbled by running the garbling scheme $G$.
In fact, the algorithm $\Pi$ can represent the instruction set for a processor circuit; hence, $\mathcal{F}$ can be realized in practice by garbling the entire processor circuit and its instruction set cf.~\cite{songhori2016garbledcpu}. 
Note the difference between our goal, i.e., realizing $\mathcal{F}$, and one achieved in~\cite{wang2016secure} is optimizing the emulation of an entire public MIPS program. 
To accomplish our goal and demonstrate a proof-of-concept, we apply available garbling tools proposed in the literature. 
More concretely, our implementation of the Garbled EDA constitutes a framework with a tool-chain to convert IP descriptions written in C, Verilog, and Assembly format to a MIPS or ARM architecture based input for an HDL-level synthesis tool.

\begin{figure}[t]
\centering \noindent
\includegraphics[width=0.9\columnwidth]{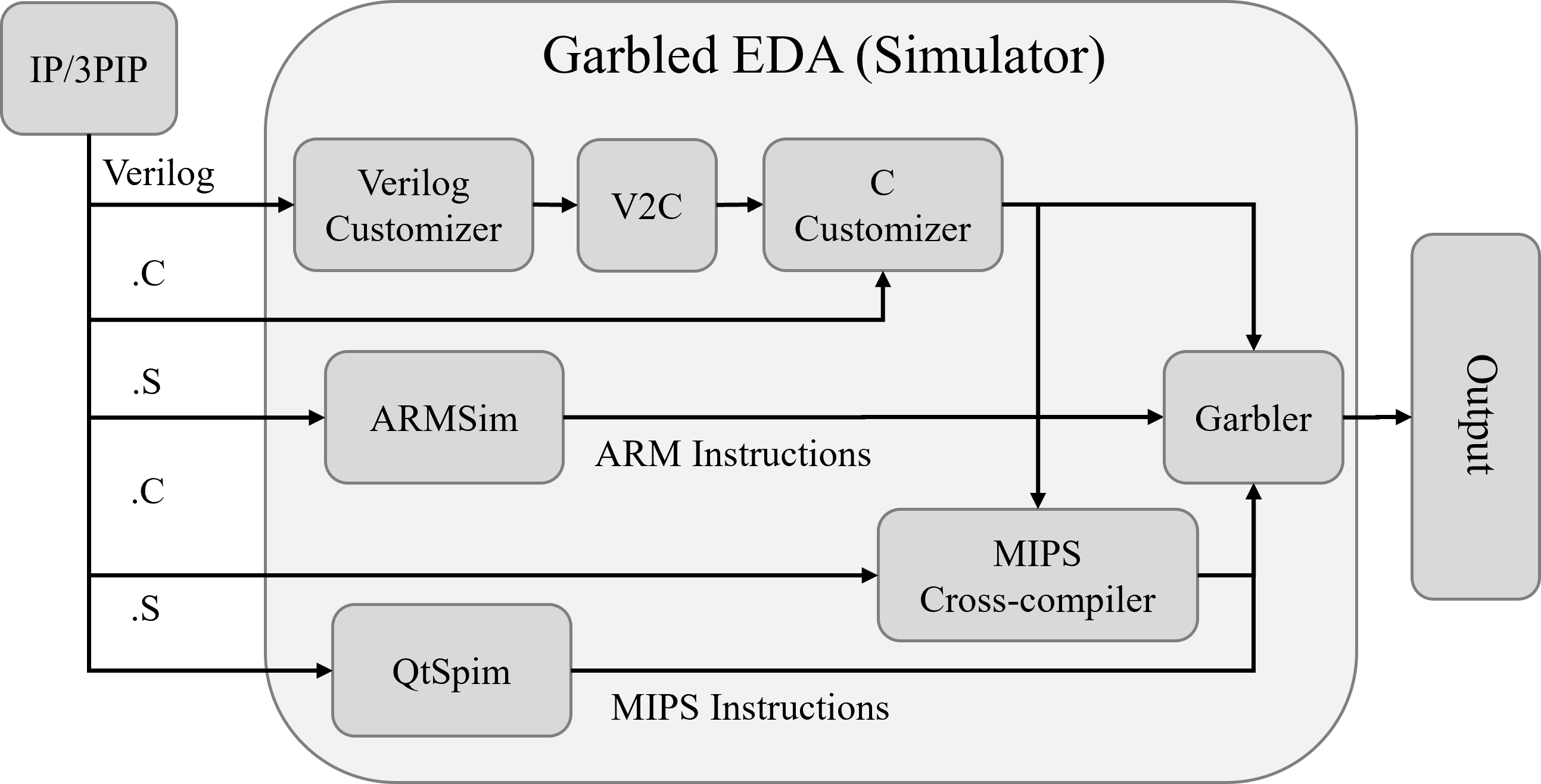}
\caption {General flow of Garbled EDA. \vspace{-5pt} }
\label{Garbled_Compiler_Flow}
\end{figure}

Although the flow introduced above is the same for simulation and compilation, we have focused on simulation in this paper as a proof-of-concept. 
Following our methodology, users can simulate an IP in a hostile environment without revealing any information about the IP or inputs.
Garbled EDA uses either MIPS or ARM architecture processors to simulate the given IP description. 
The flow of the Garbled EDA is related to the IP description format and which architecture it uses to simulate the given IP description.
Figure~\ref{Garbled_Compiler_Flow} shows the general flow of generating Garbled EDA where the garbler consists of ARM2GC framework~\cite{songhori2019arm2gc} as ARM garbler and GarbledCPU framework~\cite{songhori2016garbledcpu} as MIPS garbler.

\begin{figure*}[t!]
\centering \noindent
\includegraphics[width=\textwidth]{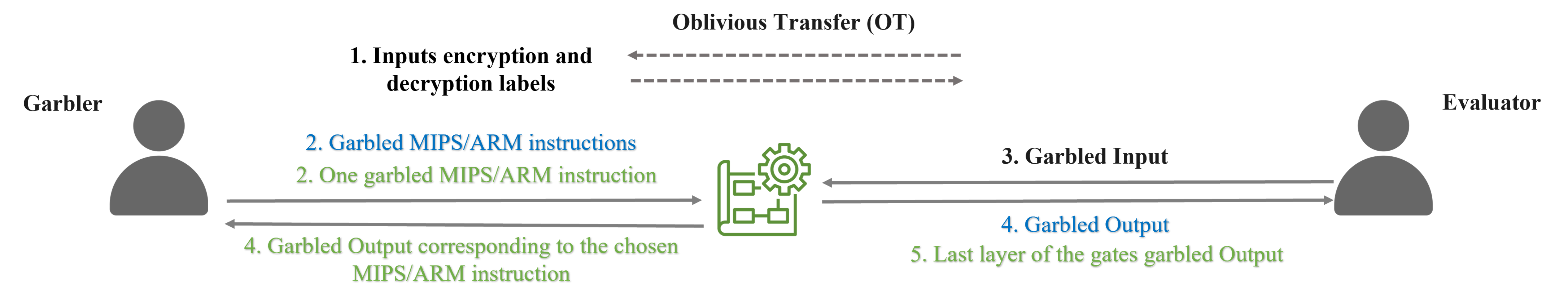}
\caption {Garbled EDA simulator with (a) the maximum performance (blue), (b) the improved resource efficiency (green).\vspace{-5pt}}
\label{Garbled EDA_flow}
\end{figure*}

If the IP description file is presented in Verilog format, the Garbled EDA simulates the IP in four steps.
First, Garbled EDA parses the IP Verilog description to find out how the IP is presented.
To give an example, an AND gate, with A and B as its inputs and C as its output, can be described as \texttt{C = A \& B;}, and  \texttt{Modulename (C,A,B);}, or \texttt{C = AND(A,B);}.
As the Garbled EDA uses the V2C tool \cite{v2c} to generate C code equivalent to the IP Verilog description and the only acceptable description format for the V2C input file is \texttt{C = A \& B;}, the Verilog Customizer part of Garbled EDA parses the IP description and converts it to a custom Verilog file suitable for the V2C tool.
In the next step, V2C generates a C file from a customized Verilog file.
Then, based on which architecture processor Garbled EDA uses to simulate the IP description, this step is divided into two parts.
If the Garbled EDA uses ARM architecture processor to simulate IP description, the C customizer maps the inputs of IP description to two memories, G and E, as defined in ARM2GC, calls the top function of IP description in gc\_main, and maps the outputs of IP to O memory; then, garbler (ARM2GC) generates garbled ARM instructions corresponding to IP functionality that executes garbled ARM instructions using TinyGarble~\cite{songhori2015tinygarble} framework.
If the Garbled EDA uses MIPS architecture processor to simulate IP description, MIPS cross-compiler compiles C file generated by V2C; thereafter, garbler (GarbledCPU) generates garbled MIPS instructions and executes them using  TinyGarble~\cite{songhori2015tinygarble}.

In the case of an IP description in C format, the Garbled EDA simulates the given IP description in two steps.
In the case that Garbled EDA uses ARM architecture processor to simulate the given IP description, the C customizer maps IP inputs to E and G memories, calls the IP top function in gc\_main, and maps IP description outputs to O memory; then, garbler (ARM2GC) generates garbled ARM instructions corresponding to IP functionality and executes them using TinyGarble.
If the IP description is presented in Assembly format, Garbled EDA simulates the given IP description in two steps.
First, Garbled EDA uses ARMSim~\cite{ARMSim} to generate ARM instructions or QtSpim~\cite{QtSpim} to generate MIPS instructions from an assembly code.
Then, garbler generates garbled ARM/MIPS instructions. 

The flow of the Garbled EDA simulator is illustrated in Figure~\ref{Garbled EDA_flow}.
The blue steps correspond to the maximum performance and the green steps correspond to the improved hardware resource efficiency implementation flow of Garbled EDA.
The rest of the steps are common in both implementation flows.

\vspace{5pt}\noindent\textbf{Garbled EDA simulator implementation flow with the maximum performance:} First, Garbler generates the MIPS/ARM instructions corresponding to the function and a set of 1 and 0 encryption and decryption labels.
In the next step, the encryption and decryption labels are sent to Evaluator through OT and the garbled MIPS/ARM instructions are fed to the garbled MIPS/ARM evaluator core.
Thereafter, Evaluator generates its garbled input corresponding to the 1 and 0 encryption labels and provides MIPS/ARM evaluator core with its garbled input.
In the last step, Evaluator receives its garbled output from the MIPS/ARM evaluator core and decrypts them using the 1 and 0 decryption labels.

As this setup requires only one OT interaction, and OT interaction costs much more time compared to the evaluation phase, this setup represents the maximum performance mode of Garbled EDA (refer to Section~\ref{sec:Garbled_Simulator_Evaluation} for more information).

\vspace{5pt}\noindent\textbf{Garbled EDA simulator implementation flow with an improved hardware resource efficiency: }
The idea of dividing the netlist into separate sub-netlists and evaluating each of them individually to decrease the memory utilization costs has been introduced in \cite{hussain2020tinygarble2}.
In our approach, we utilized the same concept and evaluate any garbled netlist by evaluating its sub-netlists.
A sub-netlist contains some neighbor gates in the main netlist and its number of gates can be chosen between the range of only one gate up to the complete netlist gate size.
In the flow of Garbled EDA with an improved hardware resource utilization, we chose the size of all sub-netlists equal to only one gate.
This means that for a netlist with $N$ gates, we have $N$ sub-netlists each having two inputs and one output. Based on Figure~\ref{Garbled EDA_flow} (green steps), in the first step, Garbler generates the garbled MIPS/ARM instruction corresponding to each gate individually and a set of 1 and 0 encryption and decryption labels.
In the next step (step 2), Garbler sends the encryption labels to the Evaluator through OT and only one garbled MIPS/ARM instruction to the garbled MIPS/ARM evaluator core.
In step 3, Evaluator generates two garbled inputs corresponding to the chosen gate by Garbler, which is unknown to Evaluator and feeds them to the garbled MIPS/ARM evaluator core.
In step 4, the generated garbled output is sent back to Garbler and based on the evaluation map (which is only held by Garbler) the next garbled MIPS/ARM instruction corresponding to the next gate of the evaluation order is sent to the core.
Steps 2, 3, and 4 repeat until all the gates that its input is connected to in the netlist are evaluated.
Then, the garbled MIPS/ARM instructions corresponding to the rest of the gates, which neither their inputs nor their output are connected to the netlist inputs and output, respectively, along with the middle wire labels are sent to the garbled MIPS/ARM evaluator core one by one until only the gates that are connected to the output of the netlist remain unevaluated.
In the last step, Garbler sends garbled MIPS/ARM instruction and Garbled input to the garbled MIPS/ARM evaluator core, and output decryption labels to Evaluator through OT.
Finally, after all the gates connected to the netlist output are evaluated, Evaluator decrypts the garbled output.

Due to the nature of the garbled circuit protocol, each gate is converted to a garbled table, and to evaluate this garbled table, only a select operation is needed \cite{bellare2012foundations}.
Hence, each garbled MIPS/ARM instruction covers four select operations which means the evaluation of each garbled MIPS/ARM instruction equals the evaluation of four garbled tables (four netlist gates).
Compared to the Garbled EDA flow with the maximum performance, this implementation flow of Garbled EDA requires more OT interactions.
The number of OT interactions required for the implementation of Garbled EDA with the hardware resource efficiency equals $M=N_{gates}/4 + (I_{size}+O_{size})$ where $N_{gates}$ is the number of the gates in the netlist, $I_{size}$ is the netlist input size, and $O_{size}$ equals to the netlist output size.
The increment of the number of OT interactions decreases the performance of the Garbled EDA but also decreases the hardware resource utilization of garbled MIPS/ARM evaluator core implementation to a constant amount, equal to a core that can execute only one instruction per cycle.

\vspace{5pt}\noindent\textbf{Garbled EDA simulator implementation flow with an improved hardware resource efficiency in the presence of a malicious adversary: }
Although Hussain et al.~\cite{hussain2020tinygarble2} have put forward an efficient way to tackle the issue with malicious adversaries, their implementation is software-based, i.e., no hardware implementation of their approach has been provided so far, to the best of our knowledge. 
Our study proposes such an implementation, as explained before.  
More specifically, the evaluation order of the garbled MIPS/ARM evaluator core is only held by Garbler.
This means the evaluation of the sub-netlists can be done in a randomized manner, known only by Evaluator.
The only evaluation order that must be fixed is for the gates connected to the output of the netlist; otherwise, Evaluator cannot decrypt the garbled output correctly.
Hence, even if the malicious adversary takes all his ability into account, to decrypt the received instruction, and middle wire labels, the only information she can get, in the best case, is the information of four gates (corresponding to the received instruction) randomly out of all netlist gates, size of the input, and the output gates.
To the best of our knowledge, there is no known attack with the ability to reveal the netlist based on this information.

%% file: Garbled_Compiler_Evaluation.tex
\section{Garbled EDA Evaluation}\label{sec:Garbled_Simulator_Evaluation}


\subsection{Evaluation Setup}\label{Evaluation_setup}
As mentioned in Section~\ref{sec:Garbled_Simulator}, an HDL-level synthesis tool can synthesize and deliver the output of Garbled EDA.
To implement garbled IP-specific simulators for each benchmark based on ARM  and MIPS architectures, we use the modified cores of Amber and Plasma processors, respectively~\cite{songhori2019arm2gc,songhori2016garbledcpu}, that can execute the garbled instructions corresponding to the given IP description. 

To examine the possibility of synthesizing and implementing IP-specific simulators, we generate the garbled ARM/MIPS instructions corresponding to different IP descriptions, store them in the instruction memory of garbled IP-specific simulators, and
use Vivado 2021 to synthesize them.
We use Vivado 2021 as well on an ARTIX-7 FPGA to evaluate the generated garbled IP-specific simulators of different IP descriptions. 
We use the ISCAS-85 benchmark suite, which is a netlist set of industrial designs. 
As their high-level design is unknown, they are good candidates for random logic circuits.
The original ISCAS-85 set consists of C432, C499, C1355, C1908, C3540, and C6288 benchmarks.
Moreover, we evaluate the Garbled EDA implementation cost of ISCAS-85 benchmarks in SFE and compare them with the Garbled EDA implementation cost of well-known benchmarks such as AES, 8-bit SUM, Hamming distance, and 8/256-bit MULT.


\begin{table}[t!]
\scriptsize
    \setlength{\tabcolsep}{0.5em}
    \renewcommand{\arraystretch}{1.5}
        \begin{center}
        \caption{Garbled EDA with maximum performance implementation cost of different benchmarks in ARM(MIPS).\label{Tab_1}}
            \begin{tabular}{|c|c|c|c|c|c|c|}
                \hline
                \textbf{Benchmark}&\textbf{XOR}&\textbf{Other}&\textbf{Inst.}& \textbf{LUT}&\textbf{Register}&\textbf{MUX}\\
                \hline
                Amber(Plasma)&N/A&N/A&64&3526 (1817)&1830 (1255)&229 (292)\\ 
                \hline
                8-bit SUM&48&96&30&22796 (9319)&20169 (7166)&1672 (523)\\
                \hline
                16-bit Hamming&3&39&47&25116 (17591)&21835 (13742)&1646 (1179)\\
                \hline
                8-bit MULT&43&139&93&32075 (32973)&26262 (25598)&1840 (1864)\\
                \hline
                C499&104&198&236&86445 (110012)&58582 (83278)&4016 (9105)\\ 
                \hline
                C432&18&222&276&91829 (130163)&69104 (97263)&5487 (11390)\\ 
                \hline
                AES&2112&576&426&158870 (287334)&102112 (200179)&6896 (16085)\\ 
                \hline
                C1355&0&746&1335&237967 (506471)&145800 (394247)&9816 (26025)\\ 
                \hline
                C1908&0&880&1560&271495 (528364)&167340 (417282)&10911 (27752)\\ 
                \hline
                256-bit MULT&1303&1276&2012&361637 (712398)&220591 (512895)&14856 (39063)\\ 
                \hline
                C3540&0&1669&3008&513105 (1009021)&306512 (787454)&19792 (52407)\\
                \hline
                C6288&0&2416&4669&788181 (1523564)&465723 (944478)&30080 (60445)\\ 
                \hline
            \end{tabular}
    \end{center}
    \vspace{-12pt}
\end{table}

\subsection{Resource Utilization Evaluation}\label{Garbled_EDA_Evaluation}
\noindent\textbf{Garbled EDA with the maximum performance hardware resource evaluation: }
Table~\ref{Tab_1} shows the Garbled EDA with the maximum performance hardware resource implementation cost of benchmarks presented in Section~\ref{Evaluation_setup}, and original Amber~\cite{Amber} and Plasma~\cite{Plasma} cores.
In Table~\ref{Tab_1}, the implementation cost using ARM (MIPS) architecture is shown outside (inside) brackets in the last three columns. 
Also, all Garbled EDA using ARM and MIPS architectures utilize 3 and 0 digital signal processing (DSP) blocks, respectively.
Moreover, we use the Garbled EDA gate parser to calculate the number of XOR (XOR column) and non-XOR gates (Other column).
The Inst. column contains the number of instructions corresponding to a cycle of loading inputs from data memory, executing the simulation, and writing the output to data memory.

As Amber and Plasma cores are two open-source cores implemented on FPGA, only their FPGA resources cost are available. 
Hence, the number of gates does not apply to them.
We include these two original cores to provide an estimate of Garbled EDA overhead.
Amber and Plasma benchmarks are the original (not garbled) cores, that can execute the ARM and MIPS instructions, respectively.
Both MIPS and ARM architectures execute an instruction in each cycle of clock~\cite{songhori2015tinygarble},~\cite{songhori2019arm2gc}.
Therefore, the execution time requirement of each Garbled EDA to calculate the output corresponding to each input is equal to the number of instructions multiplied by the FPGA clock period in nanoseconds (ns).
\textit{Hence, one of the advantages of simulating an IP using the garbling technique with Garbler EDA is that the execution time of the Garbled EDA and the MIPS/ARM format of a benchmark is the same.}

Another noticeable point of Table~\ref{Tab_1} is that using Free-XOR optimization sightly decreases the implementation cost of Garbled EDA.
To be more specific, AES benchmark has only 272 more gates than C6288, but Garbled EDA of C6288 implementation utilizes up to four times more logical/memory resources. Also, Garbled EDA of AES instruction set number is almost half of the number of the C6288 garbled instruction set.
In another case, although the 256-bit multiplication benchmark has 163 more gates than C6288, its Garbled EDA implementation utilizes about half of the C6288 Garbled EDA implementation logical and memory resources cost due to the Free-XOR optimization.
To sum it up, the more XOR gates in a benchmark, the less its Garbled EDA utilizes logical and memory resources, even if their total number of gates is equal.

In our work, we consider Garbled EDA implementation of worst-case benchmarks because we include C1355, C1908, C3540, and C6288, which contain no XOR gates, to report the maximum resource utilization overhead cost of simulating an IP with Garbled EDA. 
It should be noted that the Garbled EDA implementation overhead cost of a benchmark without any XOR gate in the PFE standard is tremendous.
On average, Garbled EDA of a benchmark utilizes logical resources $8.5\times$ and $9.5\times$ more  compared to their ARM and MIPS based architecture designs are, respectively.
Also, on average, Garbled EDA of a benchmark utilizes $10.5\times$ and $8\times$ more memory resources compared to the benchmark on ARM and MIPS based architecture design, respectively.
However, Garbled EDA of a benchmark in the PFE scheme can simulate the benchmark with the highest level of security compared to the SFE or public function evaluation.
It should be noted that the implementation cost of Amber (Plasma) benchmarks in Table~\ref{Tab_1} is for 64 instructions.
However, the more instructions, the greater the implementation costs even for this baseline.

Another important point in Table~\ref{Tab_1} is the comparison between Garbled EDA implemented in ARM and MIPS architectures.
One of the differences between these architectures is the utilization of DSP blocks.
Based on the application, if the number of DPS blocks is limited or no DSP block is available, one can use MIPS architecture.
However, as can be seen in Table~\ref{Tab_1}, simulators based on MIPS architecture are less resource-intensive than ARM-based ones when the number of instructions is less than $\approx$90; otherwise, ARM is more resource-efficient. 
Therefore, one can generate a Garbled EDA of a larger benchmark using ARM architecture in the applications with resource limitations if the DSP blocks are available.

It is obvious that both ARM and MIPS simulators impose implementation overhead to the IP designs.
As shown in Table~\ref{Tab_1}, the FPGA implementation of Garbled EDA offering the maximum performance utilizes more hardware resource utilization.
To overcome this, next, we have presented two hardware resource optimization methods for Garbled EDA.

\vspace{5pt}\noindent\textbf{Garbled EDA with a selector:} The idea of the Garbled EDA with a selector is to combine different IPs to decrease the implementation costs of their Garbled EDA compared to implementing Garbled EDA of each of them individually.
By combining different IPs and generate a Garbled EDA with a selector, implementation cost of the protocol and evaluator core is divided by the number of included IPs.
However, the instruction memory, labels, and garbled tables are multiplied by the number of included IPs.
As an example of practical implementation of the Garbled EDA with a selector, we combine C499, C432, and AES and generate a Garbled EDA with a selector corresponding to them. 
Table~\ref{Tab_2} shows the resource utilization of Garbled EDA of C499, C432, and AES compared to Garbled EDA with a selector.
\begin{table}[t!]
\scriptsize
    \setlength{\tabcolsep}{0.5em}
    \renewcommand{\arraystretch}{1.5}
        \begin{center}
        \caption{Comparison between implementation costs of Garbled EDA (maximum performance) with a selector vs. Garbled EDA of individual benchmarks.\label{Tab_2}}
            \begin{tabular}{|c|c|c|c|c|c|c|}
                \hline
                \textbf{Benchmark}&\textbf{XOR}&\textbf{Other}&\textbf{Inst.}& \textbf{LUT}&\textbf{Register}&\textbf{MUX}\\ 
                \hline
                C499&104&198&236&86445 (110012)&58582 (83278)&4016 (9105)\\ 
                \hline
                C432&18&222&276&91829 (130163)&69104 (97263)&5487 (11390)\\ 
                \hline
                AES&2112&576&426&158870 (287334)&102112 (200179)&6896 (16085)\\ 
                \hline
                \textbf{Combination}&\textbf{2366}&\textbf{1053}&\textbf{491}&\textbf{176134 (290873)}&\textbf{264937 (417391)}&\textbf{4608 (14936)}\\ 
                \hline
            \end{tabular}
    \end{center}
        \vspace{-10pt}
\end{table}
Based on Table~\ref{Tab_2}, the Garbled EDA with a selector utilizes 47.3\% (44.8\%) fewer LUTs, and 78.9\% (59.2\%) fewer MUXes, compared to the total implementation cost of Garbled EDA of each of them individually, due to the discard of two evaluation cores.
However, it utilizes 13.3\% (8.79\%) more registers in ARM (MIPS) architectures. This is because all instructions of three benchmarks are stored in instruction memory (which is plentiful), rather than wasting logical resources.

\vspace{5pt}\noindent\textbf{Garbled EDA with an improved hardware resource efficiency evaluation: }
As the Garbled EDA with an improved hardware resource efficiency implementation flow evaluates only one instruction per cycle, the Garbled MIPS/ARM evaluator core used in this implementation flow is fixed.
Hence, the hardware resource utilization of the Garbled MIPS/ARM evaluator is independent of the design.
The only difference between the evaluation of different benchmarks is the number of OT interactions.
Table~\ref{Tab_4} contains the hardware resource utilization of the implementation of a small benchmark (8-bit SUM), a well-known cryptography module (AES), a moderate size benchmark (256-bit MULT), and a huge benchmark (C6288).
As observable in Table~\ref{Tab_4}, for all the above-mentioned benchmarks, the hardware resource utilization of Garbled EDA with an improved resource efficiency is fixed.
Compared to the Garbled EDA with maximum performance implementation hardware resource utilization costs, Garbled EDA with an improved hardware resource efficiency utilizes up to $11\times$($5\times$), $16\times$($18\times$), and $406\times$($858\times$) less hardware logical resource and $24\times$($5\times$), $31\times$($18\times$), and $565\times$($671\times$) less memory resource utilization for small, moderate, and huge benchmarks, respectively, on average, where the amount inside the parenthesis are corresponding to implementation flow using Garbled MIPS evaluator core and outside parenthesis are corresponding to Garbled ARM evaluator core.
Based on Table~\ref{Tab_4} results, also mentioned by \cite{hussain2020tinygarble2}, the flow of Garbled EDA with the improved hardware resource efficiency implementation is suitable for the benchmarks that their input/output sizes are negligible compared to the number of gates.
However, even for the small benchmarks, such as 8-bit SUM, Garbled EDA with the improved hardware resource efficiency introduces a high degree of hardware resource utilization savings.
\begin{table}[t!]
\scriptsize
    \setlength{\tabcolsep}{0.5em}
    \renewcommand{\arraystretch}{1.5}
        \begin{center}
        \caption{Garbled EDA with an improved hardware resource efficiency implementation cost of different benchmarks in ARM(MIPS).\label{Tab_4}}
            \begin{tabular}{|c|c|c|c|c|c|c|c|}
                \hline
                \textbf{Benchmark}&\textbf{XOR}&\textbf{Other}&\textbf{Inst.}& \textbf{LUT}&\textbf{Register}&\textbf{MUX}&\textbf{OT}\\ 
                \hline
                8-bit SUM &48&96 &30&\multirow{4}{*}{1941(1775)}&\multirow{4}{*}{823(1406)}&\multirow{4}{*}{152(289)}&64\\ 
                \cline{1-4}\cline{8-8}
                AES&2112&576&426&&&&682\\ 
                \cline{1-4}\cline{8-8}
                256-bit MULT&1303&1276&2012&&&&3036\\ 
                \cline{1-4}\cline{8-8}
                C6288&0&2416&4669&&&&4733\\ 
                \hline
            \end{tabular}
    \end{center}
        \vspace{-10pt}
\end{table}

\vspace{5pt}\noindent\textbf{Garbled EDA execution time and peak memory cost evaluation: }
To evaluate Garbled EDA in both maximum performance and resource-efficient modes, we used a machine with Intel
454 Core i7-7700 CPU @ 3.60GHz, 16 GBs RAM, and Linux Ubuntu 20 as Garbler and an ARTIX7
455 FPGA board as the Garbled MIPS/ARM Evaluator core, which has a clock frequency of 20 MHz. Table~\ref{Tab_5} contains the execution time and peak memory for both of Garbled EDA's implementation flows (maximum performance and resource efficient).
\begin{table}[t!]
\scriptsize
    \setlength{\tabcolsep}{0.5em}
    \renewcommand{\arraystretch}{1.5}
        \begin{center}
        \caption{Comparison between implementation costs of Garbled EDA (maximum performance) vs. Garbled EDA (resource efficient) for small, moderate, and huge benchmarks.\label{Tab_5}}
            \begin{tabular}{|c|c|c|c|c|c|c|}
                \hline
                \multirow{2}{*}{\textbf{Benchmark}}&\multirow{2}{*}{\textbf{Inst.}}& \multirow{2}{*}{\makecell{\textbf{OT}\\\textbf{(Resource}\\\textbf{Efficient)}}}&\multicolumn{2}{c|}{\textbf{Time (Sec)}}&\multicolumn{2}{c|}{\textbf{Peak Memory (MB)}}\\ 
                \cline{4-7}
                &&&\makecell{Maximum\\Performance}&\makecell{Resource\\Efficient}&\makecell{Maximum\\Performance}&\makecell{Resource\\Efficient}\\
                \hline
                8-bit SUM&30&64&4.9E-5&3E-3&6.8&0.33\\ 
                \hline
                AES&426&682&6.2E-5&1.6E-2&51.2&3.54\\ 
                \hline
                256-bit MULT&2012&3036&1E-4&7.3E-2&102.4&15.12\\ 
                \hline
                C6288&4669&4733&2.3E-4&1.1E-1&25.6&23.66\\ 
                \hline
            \end{tabular}
    \end{center}
        \vspace{-10pt}
\end{table}
In the case of evaluation of Garbled EDA with an improved hardware resource efficiency, both Garbled MIPS and Garbled ARM evaluator core evaluate only one instruction per cycle, and their instruction size and number of instructions are equal, their execution times and peak memory costs are the same.
Hence we report them in one column.
Moreover, the same rule is applicable for the maximum performance case.
Since both Garbled MIPS and ARM evaluator core are implemented in a fully unrolled manner, their instructions are equal, and their input sizes are the same, the execution time and peak memory cost for both of them are identical.
The number of OT interactions for resource-efficient implementation equals the number of instructions plus the accumulation of size of netlist input and output \cite{hussain2020tinygarble2} while the number of OT interactions for maximum performance implementation equals the summation of the size of input and output.
In maximum performance implementation, each instruction takes 50 nanoseconds (ns), one design clock cycle, to be executed while in the resource-efficient implementation it takes 150ns (three design clock cycles).
Moreover, each OT interaction cost 24 microseconds (us) to be completed in both cases.
For both implementations, each OT interaction requires 3.2 kilobytes (kB) of memory for each input/output bit.
For the resource-efficient implementation, since the size of instructions and middle wire labels are small (only one instruction per and two labels per gate), for each gate, the resource-efficient implementation introduces a 2~kB memory cost. 
In the maximum performance implementation, since the encryption and decryption labels must be sent to the Evaluator along with all instructions and middle wire labels to the Garbled MIPS/ARM evaluator core at once, the maximum performance implementation requires more peak memory compared to the resource-efficient implementation in general.

%% file: Conclusion.tex
\section{Conclusion}\label{sec:conclution}
In this paper, we introduced Garbled EDA, a framework that allows IP owners to secure their IPs in the hands of end-users. Garbled EDA also provides end-users the ability to simulate, synthesize, and implement the IP design without revealing any information about the IP functionality or proprietary inputs. The introduction of a selector alleviates the tremendous logical resource utilization cost of IP design implementation up to five times with a negligible overhead in terms of memory utilization. 
Moreover, Garbled EDA with an improved hardware resource utilization reduces the cost of implementation, although there is a trade-off between performance and cost. Compared to the IEEE P1735, logic locking, and other ad hoc solutions, Garbled EDA resolves numerous problems (e.g., trusted EDA vendor) and is provably secure. In future work, we hope to demonstrate Garbled EDA for IP-specific compilers and security tools as well as consider SFE/PFE for FPGA application scenarios. Garbled EDA could be an excellent way to protect the source code/binaries and output of specialized security assessment EDA tools, thereby preventing them from being misused by attackers to find vulnerabilities in IPs and ICs.

\balance